\documentclass[letterpaper, 10 pt, conference]{ieeeconf}  % Comment this line out
\IEEEoverridecommandlockouts                              % This command is only
\usepackage{graphics} % for pdf, bitmapped graphics files
\usepackage{svg}
\usepackage{epsfig} % for postscript graphics files
\usepackage{times} % assumes new font selection scheme installed
\usepackage{amsmath} % assumes amsmath package installed
\usepackage{amssymb}  % assumes amsmath package installed

\usepackage{subfiles}
\usepackage{cite}
\renewcommand{\epsilon}{\varepsilon}

\newtheorem{theorem}{Theorem}

\newtheorem{corollary}{Corollary}
\newtheorem{lemma}{Lemma}
\newtheorem{assumption}{Assumption}
\newtheorem{definition}{Definition}

\newtheorem{remark}{Remark}

\renewcommand{\epsilon}{\varepsilon}

\newcommand{\lf}{\left}
\newcommand{\rg}{\right}

\newcommand{\ii}{\infty}

\newcommand{\dee}{\,\mathrm{d}}

\newcommand{\argmin}{\operatorname{arg\,min}}
\newcommand{\argmax}{\operatorname{arg\,max}}

\newcommand{\sbs}{\subseteq}

\newcommand{\ssbs}{\subset}

\newcommand{\tms}{\times}

\newcommand{\ha}{\hat}

\newcommand{\R}{\mathbb{R}}

\newcommand{\al}{\alpha}
\newcommand{\bt}{\beta}

\newcommand{\dl}{\delta}

\newcommand{\ep}{\epsilon}

\newcommand{\et}{\eta}

\newcommand{\kp}{\kappa}
\newcommand{\lm}{\lambda}
\newcommand{\Lm}{\Lambda}

\newcommand{\ta}{\tau}

\newcommand{\Ru}{\mathbb{R}^{n_\textup{u}}}
\newcommand{\Rd}{\mathbb{R}^{n_\textup{d}}}
\newcommand{\Ry}{\mathbb{R}^{n_\textup{y}}}
\newcommand{\Rz}{\mathbb{R}^{n_\textup{z}}}
\newcommand{\Rzy}{\mathbb{R}^{n_\textup{z} \times n_\textup{y}}}
\newcommand{\Ryy}{\mathbb{R}^{n_\textup{y} \times n_\textup{y}}}

\newcommand{\tVal}{t}
\newcommand{\uVal}{u}
\newcommand{\dVal}{d}
\newcommand{\zVal}{z}
\newcommand{\yVal}{y}
\newcommand{\wVal}{w}

\newcommand{\uVals}{\mathcal{U}}
\newcommand{\dVals}{\mathcal{D}}

\newcommand{\uSig}{\mathbf{u}}
\newcommand{\dSig}{\mathbf{d}}
\newcommand{\zSig}{\mathbf{z}}
\newcommand{\ySig}{\mathbf{y}}
\newcommand{\wSig}{\mathbf{w}}
\newcommand{\aSig}{\mathbf{h}}

\newcommand{\uSigs}{\mathbb{U}_{[t,0)}}
\newcommand{\dSigs}{\mathbb{D}_{[t,0)}}
\newcommand{\uSigsts}{\mathbb{U}_{[t,s)}}
\newcommand{\dSigsts}{\mathbb{D}_{[t,s)}}

\newcommand{\uSigsi}[1]{\mathbb{U}_{#1}}
\newcommand{\dSigsi}[1]{\mathbb{D}_{#1}}
\newcommand{\uSigsii}[2]{\mathbb{U}_{[t_{#1}, t_{#2})}}
\newcommand{\dSigsii}[2]{\mathbb{D}_{[t_{#1}, t_{#2})}}

\newcommand{\dStrat}{\gamma}

\newcommand{\dStrats}{\Gamma_{[t,0)}}

\newcommand{\targetSet}{\mathcal{S}}
\newcommand{\BRS}{\textup{BRS}}
\newcommand{\BRT}{\textup{BRT}}
\newcommand{\BST}{\textup{BST}}

\newcommand{\zCompact}{\mathcal{Z}}
\newcommand{\yCompact}{\mathcal{Y}}
\newcommand{\tCompact}{\mathcal{T}}

\newcommand{\trajSPz}{\mathbf{z}_{t, z, y}^{\epsilon, \mathbf{u}, \mathbf{d}}}
\newcommand{\trajSPy}{\mathbf{y}_{t, z, y}^{\epsilon, \mathbf{u}, \mathbf{d}}}
\newcommand{\trajSPzstrat}{\mathbf{z}_{t, z, y}^{\epsilon, \mathbf{u}, \gamma[\mathbf{u}]}}

\newcommand{\trajRM}{\bar{\mathbf{z}}_{t, z}^{\mathbf{u}, \mathbf{d}}}
\newcommand{\trajRMstrat}{\bar{\mathbf{z}}_{t, z}^{\mathbf{u}, \gamma[\mathbf{u}]}}

\newcommand{\trajBLM}{\tilde{\mathbf{y}}_{t, y}^{z,\mathbf{u}, \mathbf{d}}}

\newcommand{\payoff}{\ell}

\newcommand{\payoffBLM}{\tilde{J}_{{t,y}}^{z,\lambda}}

\newcommand{\valSP}{V_\epsilon}
\newcommand{\valRM}{\bar{V}}

\newcommand{\valBLMdiscretePlus}{\tilde{V}_{z,\lambda}^{P,+}}
\newcommand{\valBLMdiscreteMinus}{\tilde{V}_{z,\lambda}^{P,-}}
\newcommand{\valBLMdiscretePlusMinus}{\tilde{V}_{z,\lambda}^{P,\pm}}

\title{\LARGE \bf
Approximate Hamilton-Jacobi Reachability Analysis for a Class of Two-Timescale Systems, with Application to Biological Models
}

\author{Dylan Hirsch and Sylvia Herbert% <-this % stops a space
\thanks{This material is based upon work supported by the National Science Foundation Graduate Research Fellowship Program under Grant No. DGE-2038238. 
Any opinions, findings, and conclusions or recommendations expressed in this material are those of the author(s) and do not necessarily reflect
the views of the National Science Foundation.}% <-this % stops a space
\thanks{Research reported in this publication was supported by NIBIB of the National Institutes of Health under award number T32EB009380.} %
\thanks{Dylan Hirsch (corresponding author) and Sylvia Herbert are with the Department of Mechanical and Aerospace Engineering, University of California at San Diego, 9500 Gilman Drive MC 0411, La Jolla, CA 92093.
        {\tt\small dhirsch@ucsd.edu, sherbert@ucsd.edu.}}%
}
\begin{document} 

\maketitle
\thispagestyle{empty}
\pagestyle{empty}

%%%%%%%%%%%%%%%%%%%%%%%%%%%%%%%%%%%%%%%%%%%%%%%%%%%%%%%%%%%%%%%%%%%%%%%%%%%%%%%%
\begin{abstract}

Hamilton-Jacobi reachability (HJR) is an exciting framework used for control of safety-critical systems with nonlinear and possibly uncertain dynamics.
However, HJR suffers from the curse of dimensionality, with computation times growing exponentially in the dimension of the system state.
Many autonomous and controlled systems involve dynamics that evolve on multiple timescales, and for these systems, singular perturbation methods can be used for model reduction.
However, such methods are more challenging to apply in HJR due to the presence of an underlying differential game.
In this work, we leverage prior work on singularly perturbed differential games to identify a class of systems which can be readily reduced, and we relate these results to the quantities of interest in HJR.
We demonstrate the utility of our results on two examples involving biological systems, where dynamics fitting the identified class are frequently encountered.

\end{abstract}
%%%%%%%%%%%%%%%%%%%%%%%%%%%%%%%%%%%%%%%%%%%%%%%%%%%%%%%%%%%%%%%%%%%%%%%%%%%%%%%%

\section{INTRODUCTION}

As computers have become faster, Hamilton-Jacobi reachability (HJR) is an increasingly useful tool for analyzing and controlling safety-critical systems using the theory of differential games \cite{HJR-survey}.
In particular, HJR determines the states, known as the backward reachable set (BRS), from which a system can be guaranteed able to achieve some goal, such as reaching a target or avoiding an obstacle, despite uncertain model dynamics or presence of an adversary 
\cite{Mitchell,Margellos,Fisac}.
In the process, one also obtains a control law by which to achieve the goal.
Substantial work has been performed to extend and apply HJR to problems in domains such as aerospace, robotics, and reinforcement learning \cite{AeroHJRApplication,FaSTrack,Milan}.

The key factor limiting the more widespread use of HJR is the curse of dimensionality.
Even with modern processors, the algorithm typically becomes computationally intractable for systems of dimension 6 or higher.
This limitation produces a need for model reduction techniques that can be applied in the context of differential games.

In the control theory community, singular perturbation methods are mainstays of reduced modeling of autonomous and controlled systems, specifically those that evolve on multiple timescales \cite{Kokotovic-Survey,Kokotovic-Book}.
Although results on model reduction for singularly perturbed (SP) differential games also exist, they have yet to see widespread use, particularly in the context of reachability analyses \cite{gaitsgory}.
This gap is likely in part due to the complexity of analytically finding the proper reduced model for a given SP differential game, if the game admits such a reduced model in the first place.

\subsection{Background and prior work}

Some of the earliest work studying differential games with both fast and slow state dynamics was by Gardner and Cruz in the late 1970s, in which the authors demonstrated that naively performing model reduction via singular perturbation prior to evaluating the payoff at the Nash equilibrium gives a result different from evaluating the true payoff in the limit of increasingly separated timescales \cite{Well-Posedness-Gardner}.
In their study, the fast and slow systems were linear and the payoff was quadratic.
These authors refer to such a singularly perturbed differential game as ill-posed.
Indeed, the discovery of an ill-posed game in such a simple setting motivated the careful study of singular perturbation methods for differential games.
Research along these lines was continued by Khalil and Kokotovic soon after, demonstrating the link between well-posedness and the information structure of the game \cite{Well-Posedness-Khalil}.

Later on, the advent of viscosity solutions of Hamilton-Jacobi equations \cite{Crandall-Lions,Crandall-Lions-Evans} and the study of their relationship to differential games \cite{Evans-Diff-Games,Barron-Evans-Jensen} ignited renewed interest in the study of differential games with non-anticipative information structures, nonlinear dynamics, and non-quadratic payoff functions.
Leveraging these results, guarantees of well-posedness for these general differential games under the added complication of singularly perturbed dynamics was undertaken by Gaitsgory in \cite{gaitsgory}.
Here, Gaitsgory demonstrated that such a game could be decomposed into an ``associated fast'' differential game involving only the fast dynamics and a reduced differential game involving only the slow variables.
However, certifying well-posedness using the theory outlined in this work for a specific system is typically challenging, involving a number of assumptions that may be difficult to check.
Moreover, in practice, using the theory often requires one to explicitly obtain the value function for the associated fast game, which may be impractical.
These results have also been adapted for discrete-time settings \cite{Shi-discrete}.

Extensive further work has been pursued on linear-quadratic differential games, with and without noise, on finite and infinite intervals, and for zero-sum and non-zero-sum settings \cite{LQ1,LQ2,LQ3,LQ4}.
Additional works have proceeded for the nonlinear case by approaches rooted in weak solutions to Hamilton-Jacobi Bellman equations, but each typically involves conditions that may be challenging to verify for general systems \cite{Subbotina,Alvarez-Bardi-Ergodicity,Subbotina2006}.

Thus, in this work, we instead choose to focus on a class of nonlinear systems for which well-posedness can be certified via Gaitsgory's results, and perhaps more importantly, we demonstrate how one can use this convergence result to provide bounds on quantities of interest in HJR, such as the BRS, for such systems.

\subsection{Contributions}
In this work, we provide three main contributions.
First, we identify a class of differential games for which a reduced model can be analytically derived and rigorously justified via the results in \cite{gaitsgory}.
Second, we derive inner and outer approximations of the backward reachable sets for these systems.
Finally, we provide examples of how to apply these results in the context of models of biomolecular system models, where two-timescale dynamics arise frequently \cite{Murray}.

\section{SYSTEM DESCRIPTION}

Let $\uVals \ssbs \Ru$ and $\dVals \ssbs \Rd$ be non-empty and compact.
For each $\tVal < 0$, let $\uSigs$ and $\dSigs$ be the set of measurable functions from $[\tVal,0)$ to $\uVals$ and $[\tVal,0)$ to $\dVals$, respectively.

\subsection{Singularly perturbed system}
We consider the following SP system, parameterized by a ``small'' parameter $\ep > 0$:
\begin{align}
    \dot{\zSig} &= f(\zSig, \uSig, \dSig) + M(\zSig) A(\zSig, \uSig, \dSig)\ySig \label{eqn:sp-system-slow}\\
    \epsilon \dot{\ySig} &= g(\zSig, \uSig, \dSig) + A(\zSig, \uSig, \dSig)\ySig \label{eqn:sp-system-fast},
\end{align}
where $f:\Rz \tms \uVals \tms \dVals \to \Rz$, $g:\Rz \tms \uVals \tms \dVals \to \Ry$, $M: \Rz \to \Rzy$, and $A:\Rz \tms \uVals \tms \dVals \to \Ryy$.
In the above, $\zSig$ and $\ySig$ represent the ``slow'' and ``fast'' state variables, respectively, and $\uSig$ and $\dSig$ represent control and disturbance signals, respectively.

This formulation arises commonly in models of biological systems, where simple chemical reactions often have dynamics that are affine (or nearly affine) in the species' concentrations (i.e. $\ySig$) and occur on much faster timescales than more complex processes.
We make the following assumptions:
\begin{assumption}[Regularity]\label{assumption:regularity}
        There is some $K > 0$ such that $\|M(\zVal)\|$$ + \|A(\zVal, \uVal, \dVal)\| \le K$ and $\|f(\zVal,\uVal,\dVal)\| + \|g(\zVal,\uVal,\dVal)\| \le K(1 + \|\zVal\|)$ for all $\zVal \in \Rz$, $\uVal \in \uVals$, and $\dVal \in \dVals$.
        Moreover, $f$, $g$, $M$, and $A$ are all locally Lipschitz in $z$ and continuous on their domains.
\end{assumption}
\begin{assumption}[Stability of fast dynamics]\label{assumption:stability}
There is a symmetric, positive definite matrix $P \in \Ryy$ such that 
$$A(\zVal,\uVal,\dVal)^\top P + P A(\zVal,\uVal,\dVal) \prec 0$$
for each $\zVal \in \Rz$, $\uVal \in \uVals$, and $\dVal \in \dVals$.
\end{assumption}
\begin{assumption}[Isaacs' condition for reduced model]\label{assumption:saddle-point}
For each $\lm, \zVal \in \Rz$,
\begin{align*}
    &\min_{\uVal \in \uVals} \max_{\dVal \in \dVals} \lm^\top \lf[f(\zVal,\uVal,\dVal) - M(z) g(\zVal,\uVal,\dVal) \rg]\\ 
    &\quad\quad = \max_{\dVal \in \dVals} \min_{\uVal \in \uVals} \lm^\top \lf[f(\zVal,\uVal,\dVal) - M(z) g(\zVal,\uVal,\dVal) \rg].
\end{align*}
\end{assumption}

The reduced model to which Assumption \ref{assumption:saddle-point} is related will be introduced shortly.
In practice, these assumptions are often satisfied.
Indeed, Assumption \ref{assumption:stability} is an analog of the stability requirement of standard singular perturbation methods, and Isaacs' condition is assumed commonly in differential games.

For each $\ep > 0$, $\tVal < 0$, $\zVal \in \Rz$, $\yVal \in \Ry$, $\uSig \in \uSigs$, and $\dSig \in \dSigs$, we let $\lf(\trajSPz, \trajSPy \rg):[\tVal,0] \to \Rz \tms \Ry$ be the Carath\'{e}odory solution of \eqref{eqn:sp-system-slow}-\eqref{eqn:sp-system-fast} for which $\trajSPz(t) = \zVal$ and $\trajSPy(t) = \yVal$ (existence and uniqueness of this solution follows from Assumption \ref{assumption:regularity}; see Theorem 1.2.1 in \cite{Friedman}).

\subsection{Description of the differential game}

For each value of $\ep > 0$, we consider a differential game in which the control ``player'' $\uSig$ wishes to drive the slow variable $\zSig$ to be in some open set $\targetSet \sbs \Rz$ at time 0, and the disturbance ``player'' $\dSig$ wishes to prevent this outcome.

When formulating the game mathematically, we wish to ensure the disturbance player only has information regarding the control player's decisions up to the current time.
This restriction is codified via the following definition.
\begin{definition}
    Given $\tVal < 0$, a map $\dStrat:\uSigs \to \dSigs$ is a \textit{non-anticipative disturbance strategy on} $[\tVal,0)$ if for each $s \in [\tVal,0)$ and $u_1, u_2 \in \uSigs$, we have that $\dStrat[u_1](\ta) = \dStrat[u_2](\ta)$ for a.e. $\ta \in [\tVal,s)$ whenever $u_1(\ta) = u_2(\ta)$ for a.e. $\ta \in [\tVal,s)$.
\end{definition}

We denote the set of non-anticipative disturbance strategies on $[\tVal,0)$ by $\dStrats$.

The game proceeds as follows.
Denote by $\tVal < 0$ the initial time and by $(\zVal,\yVal)$ the initial system state.
The disturbance player selects a $\dStrat \in \dStrats$, and the control player the selects a $\uSig \in \uSigs$.
We let $\dSig = \dStrat[\uSig]$ be the resulting disturbance signal.
If $\trajSPz(0) \in \targetSet$, the control player wins the game; otherwise the disturbance player wins.

\subsection{Backward reachable set and the value function}
The first quantity of interest in HJR is the backward reachable set (BRS). For each $\ep > 0$ and $\tVal < 0$, we let
\begin{align*}
    \BRS_\ep(\tVal)= \{& (\zVal,\yVal) \in \Rz \tms \Ry \mid \\
    &\forall \dStrat \in \dStrats~ \exists \uSig \in \uSigs~ \trajSPzstrat(0) \in \targetSet\}.
\end{align*}
Note that $\BRS_\ep(\tVal)$ represents the set of initial states from which the control player will win the game corresponding to parameter $\ep$ and inital time $t$ if both players act rationally.
For convenience, we also set $\BRS_\ep(0) = \targetSet$.

In the HJR framework, we implicitly encode the target set $\targetSet$ via a function $\ell$ and then define a value function for the game, where $\ell$ serves as the terminal payoff function.
More explicitly, we choose $\ell:\Rz \to \R$ to be a bounded, Lipschitz function for which
\begin{equation*}
    \targetSet = \{\zVal \in \Rz: \ell(\zVal) < 0\}.
\end{equation*}
For each $\ep > 0$, we define the value function $\valSP: (-\ii,0] \tms \Rz \tms \Ry \to \R$ by
\begin{equation*}
    \valSP(\tVal, \zVal, \yVal) =
    \begin{cases}
        \sup_{\dStrat \in \dStrats} \inf_{\uSig \in \uSigs} \ell \lf( \trajSPzstrat(0) \rg) & \tVal < 0, \\
        \payoff(\zVal) & \tVal = 0.
    \end{cases}
\end{equation*}
It can be seen from these definitions that for each $t \le 0$,
\begin{align}
    \{(\zVal,\yVal) \in &\Rz \tms \Ry \mid \valSP(\tVal,\zVal,\yVal) < 0 \} \sbs \BRS_\ep(\tVal) \nonumber\\
    &\sbs \{(\zVal,\yVal) \in \Rz \tms \Ry \mid \valSP(\tVal,\zVal,\yVal) \le 0 \}.\label{eqn:BRS-standard-bounds}
\end{align}

\subsection{Backward reachable tubes and backward avoid tubes}
The BRS is the set of states from which the controller can ensure the system will be in the target set \textit{at} the final time.
We may also be interested in the set of states from which the controller can ensure the system will be in the target set \textit{by} or \textit{until} the final time, known as the Backward Reachable Tube (BRT) and the Backward Stay Tube (BST), respectively.

For each $\ep > 0$ and $\tVal < 0$, we let
\begin{align*}
    \BRT_\ep&(\tVal) = \{ (\zVal,\yVal) \in \Rz \tms \Ry \mid \\
    &\forall \dStrat \in \dStrats~ \exists \uSig \in \uSigs~ \exists s \in [\tVal,0]~ \trajSPzstrat(s) \in \targetSet\},\\
    \BST_\ep&(\tVal)= \{ (\zVal,\yVal) \in \Rz \tms \Ry \mid \\
    &\forall \dStrat \in \dStrats~ \exists \uSig \in \uSigs~ \forall s \in [\tVal,0]~ \trajSPzstrat(s) \in \targetSet\}.
\end{align*}
For convenience, we also set $\BRT_\ep(0) = \BST_\ep(0) = \targetSet$.
Note that for each $t \le 0$ and $s \in [t,0]$
\begin{equation}\label{eqn:BRT-BST-bound}
    \BRS_\ep(s) \sbs \BRT_\ep(t),\text{ and }
    \BST_\ep(t) \sbs \BRS_\ep(s).
\end{equation}

\subsection{Reduced model}
We consider as a candidate for approximating the SP system the following ``reduced'' model:
\begin{align}
    \dot{\zSig} = f(\zSig,\uSig,\dSig)- M(\zSig) g(\zSig,\uSig,\dSig). \label{eqn:reduced-system}
\end{align}

For each $\tVal < 0$, $\zVal \in \Rz$, $\uSig \in \uSigs$, and $\dSig \in \dSigs$ we let $\trajRM$ be the unique Carath\'{e}odory solution of \eqref{eqn:reduced-system} for which $\trajRM(\tVal) = \zVal$ (existence and uniqueness of this solution follows from Assumption \ref{assumption:regularity}; see Theorem 1.2.1 in \cite{Friedman}).
We then define the ``reduced'' value function $\valRM: (-\ii,0] \tms \Rz \to \R$ by
\begin{equation*}
    \valRM(\tVal, \zVal) =
    \begin{cases}
        \sup_{\dStrat \in \dStrats} \inf_{\uSig \in \uSigs} \ell \lf(\trajRMstrat(0) \rg) & \tVal < 0,\\
        \ell(\zVal) & \tVal = 0.
    \end{cases}
\end{equation*}

We now explain how to use the reduced value function to obtain bounds on the BRS, BRT, and BST of the SP system.

\section{RESULTS}

The main result is given by Theorem \ref{theorem:main-theorem}, which uses level sets of the reduced value function $\valRM$ to provide inner and outer approximations of $\BRS_\ep(t)$ when $\ep$ is sufficiently small.
First, a lemma (see Appendix for its proof):

\begin{lemma}\label{lemma:main-lemma}
    Suppose Assumptions \ref{assumption:regularity}-\ref{assumption:saddle-point} hold.
    Then for any $\tCompact \ssbs (-\ii,0]$, $\zCompact \ssbs \Rz$, and $\yCompact \ssbs \Ry$ all non-empty and compact,
    \begin{equation*}
        \sup_{(t,\zVal,\yVal) \in \tCompact \tms \zCompact \tms \yCompact} |\valSP(t,\zVal,\yVal) - \valRM(t,\zVal)| \overset{\ep \to 0^+}{\relbar\joinrel\relbar\joinrel\longrightarrow} 0.
    \end{equation*}
\end{lemma}
\begin{theorem}\label{theorem:main-theorem}
    Suppose Assumptions \ref{assumption:regularity}-\ref{assumption:saddle-point} hold.
    Let $\et > 0$, and let $\tCompact \ssbs (-\ii,0]$, $\zCompact \ssbs \Rz$, and $\yCompact \ssbs \Ry$ each be compact.
    There exists an $\ep_0 > 0$ such that
    \begin{align}
        \{\zVal \in \zCompact \mid \valRM(\tVal,\zVal) < & -\et \} \tms \yCompact \sbs \BRS_\ep(\tVal) \cap (\zCompact \tms \yCompact) \nonumber\\
        &\sbs\{\zVal \in \zCompact \mid \valRM(\tVal,\zVal) < +\et \} \tms \yCompact \label{eqn:main-theorem-bound}
    \end{align}
    for all $\ep \in (0,\ep_0)$ and $t \in \tCompact$.
\end{theorem}
\begin{proof}
    By Lemma 1, we can choose $\ep_0 > 0$ such that for all $\ep \in (0, \ep_0)$, $\tVal \in \tCompact$, $\zVal \in \zCompact$, and $\yVal \in \yCompact$,
    \begin{equation*}
    |\valSP(\tVal,\zVal,\yVal) - \valRM(\tVal,\zVal)| < \et.
    \end{equation*}
    It follows that for each such $\ep$, $\tVal$, $\zVal$, and $\yVal$,
    \begin{align*}
        \{\zVal \in \zCompact &\mid \valRM(\tVal,\zVal) < -\et \} \tms \yCompact\\
        &= \{(\zVal,\yVal) \in \zCompact \tms \yCompact \mid \valRM(\tVal,\zVal) < - \et \}\\
        &\sbs \{(\zVal,\yVal) \in \zCompact \tms \yCompact \mid \valSP(\tVal,\zVal,\yVal) < 0 \}\\
        &\sbs \{(\zVal,\yVal) \in \zCompact \tms \yCompact \mid \valSP(\tVal,\zVal,\yVal) \le 0 \} \\
        &\sbs \{(\zVal,\yVal) \in \zCompact \tms \yCompact \mid \valRM(\tVal,\zVal) < + \et \}\\
        &= \{\zVal \in \zCompact \mid \valRM(\tVal,\zVal) < +\et \} \tms \yCompact.
    \end{align*}
    The result then follows from \eqref{eqn:BRS-standard-bounds}.
\end{proof}

By setting $\tCompact = [\tVal,0]$, we can also use Theorem 1 and \eqref{eqn:BRT-BST-bound} to get bounds on the BRT and BST of the SP system:
\begin{corollary}
    Suppose Assumptions \ref{assumption:regularity}-\ref{assumption:saddle-point} hold.
    Let $\et > 0$ and $t \le 0$, and let $\zCompact \ssbs \Rz$, and $\yCompact \ssbs \Ry$ both be compact.
    There exists an $\ep_0 > 0$ such that for all $\ep \in (0,\ep_0)$
    \begin{align*}
        \{\zVal \in \zCompact \mid \min_{s \in [t,0]} \valRM(s,\zVal) < -\et \} \tms \yCompact \sbs \BRT_\ep(\tVal) \cap (\zCompact \tms \yCompact),
    \end{align*}
    and
    \begin{align*}
        \BST_\ep(\tVal) \cap (\zCompact \tms \yCompact) \sbs \{\zVal \in \zCompact \mid \max_{s \in [t,0]} \valRM(s,\zVal) < +\et \} \tms \yCompact.
    \end{align*}
\end{corollary}

\section{EXAMPLES}

\subsection{Genetic circuit with negative feedback}
\begin{figure}[ht!]
    \centering
    \includegraphics[width=1.0\linewidth]{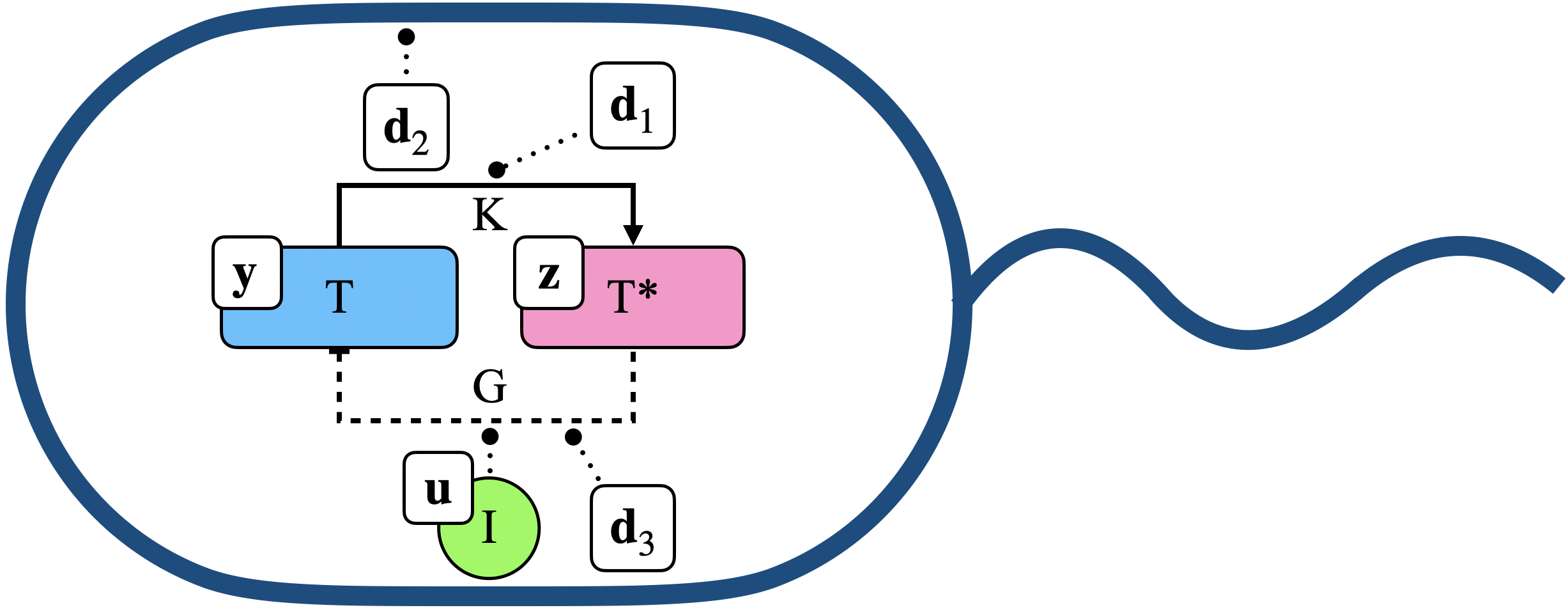}
    \caption{Depiction of a bacterium, engineered with the circuit of interest. Solid arrow represents chemical conversion of T to T* via K; dashed arrow represents down-regulation of T production by T* via binding to the promoter of G; dotted arrows represent activity modulation by control and disturbance signals. In particular, $\mathbf{u}$ (representing the inducer I) modulates the ability of T* to bind to the promoter of G, $\mathbf{d}_1$ modulates kinase activity, $\mathbf{d}_2$ modulates the cell growth rate, and $\mathbf{d}_3$ modulates the activity of G.}
    \label{fig:small-example-cartoon}
\end{figure}

We first consider a model of a simple genetic circuit engineered in a bacterium (Figure \ref{fig:small-example-cartoon}).
This circuit contains a gene G, which produces a transcription factor T.
A kinase K expressed naturally in the cell chemically modifies T into its active form T* via a reaction known as phosphorylation.
T* can then bind to the promoter region of G to down-regulate production of T, creating a negative feedback loop.
Additionally, an inhibitor molecule I can be added by the user to the cellular environment to up-regulate the production rate of T by reducing the ability of T* to bind to the promoter of G.
Because the foreign circuit pulls away machinery (e.g. RNAP, ribosomes, energy) used in other functions of the cell, quantities such as the growth rate of the cell, concentration of K, and production rate of T from G are known to fluctuate in a manner that is generally difficult to predict.
We assume the transcription factor is fluorescently tagged so that its concentration can be observed by the user.

When the rate of phosphorylation is much faster than the rate of cell growth, we can model this circuit as follows:
\begin{align}
    \dot{\zSig} &=  \al \dSig_1 \ySig - \dSig_2\zSig,\label{eqn:small-example-1}\\
    \ep \dot{\ySig} &=\dSig_3 \frac{\uSig^2}{\uSig^2 + \zSig^2} - \dSig_1 \ySig.\label{eqn:small-example-2}
\end{align}
In the above, $\zSig$ represents the (normalized) concentration of T* and $\ySig$ represents the (normalized) concentration of T, respectively.
The control $\uSig$ represents the (normalized) concentration of I, and the disturbances $\dSig_i$ capture the effects of the circuit on the natural cellular processes.
The parameter $\epsilon \ll 1$ quantifies the separation between the timescale at which the phosphorylation reaction occurs and the timescale on which cell growth, and the parameter $\alpha$ is a positive constant related to both the basal kinetic rates of the various processes and the concentrations of K and G in the cell.

The corresponding reduced model for this system is
\begin{equation}
    \dot{\zSig} = \al \dSig_3 \frac{\uSig^2}{\uSig^2 + \zSig^2} - \dSig_2\zSig.
\end{equation}

\begin{figure}[ht!]
    \centering
    \vspace*{.1 in}
    \includegraphics[width=1.0\linewidth]{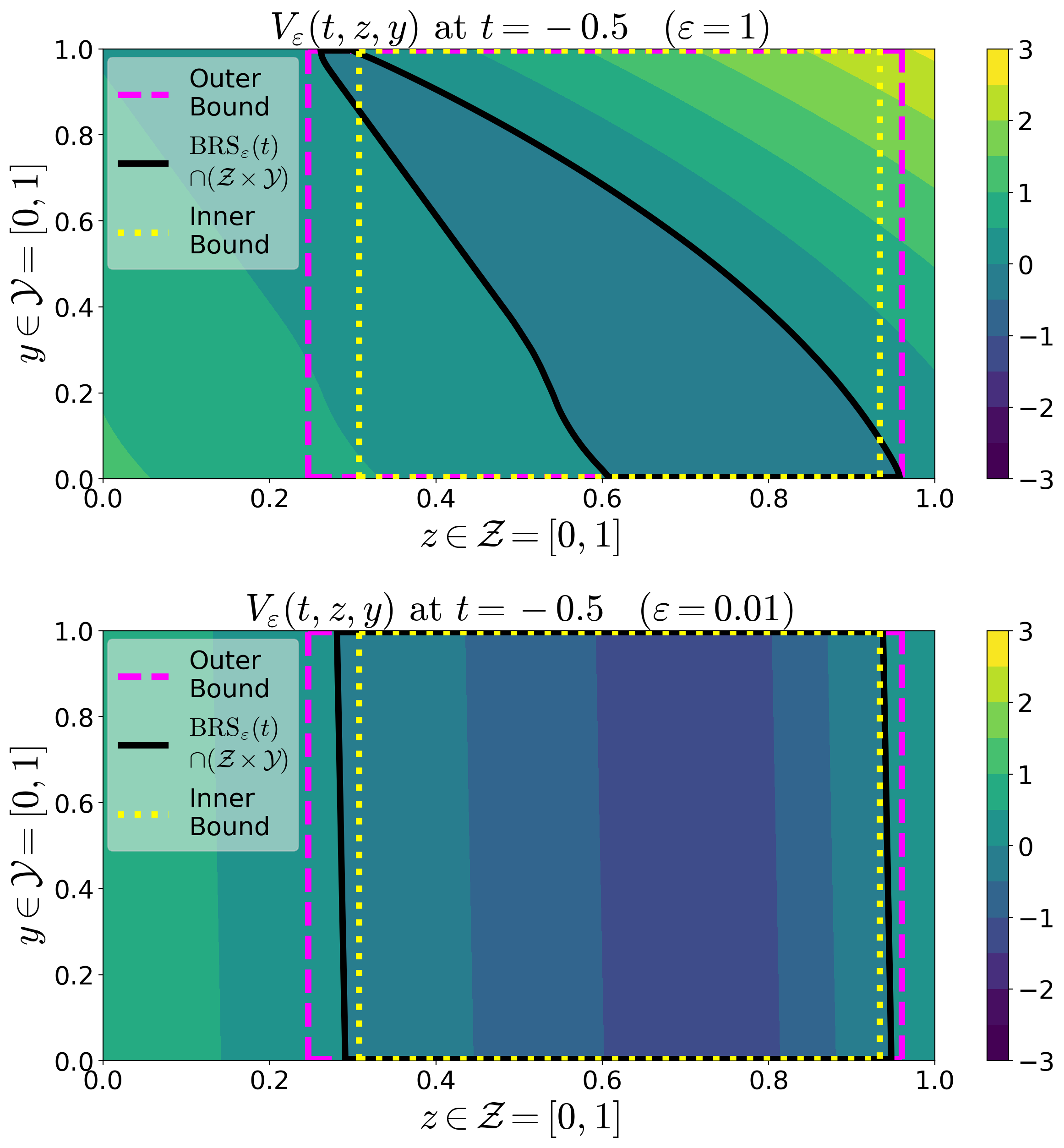}
    \caption{Contour plots at time $t = -0.5$ of the value function $V_\ep(t,z,y)$ of the SP system \eqref{eqn:small-example-1}-\eqref{eqn:small-example-2}.
    The target set is $\targetSet = (0.25,0.75)$, and the terminal payoff function is taken to be $\ell(z) = \min\{10 (|z - 0.5| - 0.25),3\}$.
    The intersection of this system's BRS with the compact set $\mathcal{Z} \tms \mathcal{Y} = [0,1]^2$ is the region inside the black curve. 
    The inner and outer bounds from \eqref{eqn:main-theorem-bound}, with $\eta = 0.1$, are the regions inside the yellow and magenta curves, respectively.
    Parameters were $\uVals = [0.1,1]$, $\dVals = [0.5,2]^3$, $\al = 1$. (Top) Results for $\ep = 1$.
    Note that the bounds do not hold because $\ep$ is not sufficiently small.
    (Bottom) Results for $\ep = 0.01$.
    Note that the bounds now do hold as $\ep$ is sufficiently small, so the system is approximately one-dimensional.
    }
    \label{fig:small-example-level-sets}
\end{figure}

The BRS of the SP model \eqref{eqn:small-example-1}-\eqref{eqn:small-example-2}, along with the inner and outer bounds from Theorem \ref{theorem:main-theorem} are shown in Figure \ref{fig:small-example-level-sets} for large and small values of $\ep$.
Note that when $\ep$ is not sufficiently small, the bounds do not necessarily hold.
However, when $\ep$ is sufficiently small, the bounds hold and the value function appears almost independent of $y$.

\subsection{Metabolic reaction network in a growing population}
\begin{figure}[ht!]
    \centering
    \vspace*{.1 in}
    \includegraphics[width=1.0\linewidth]{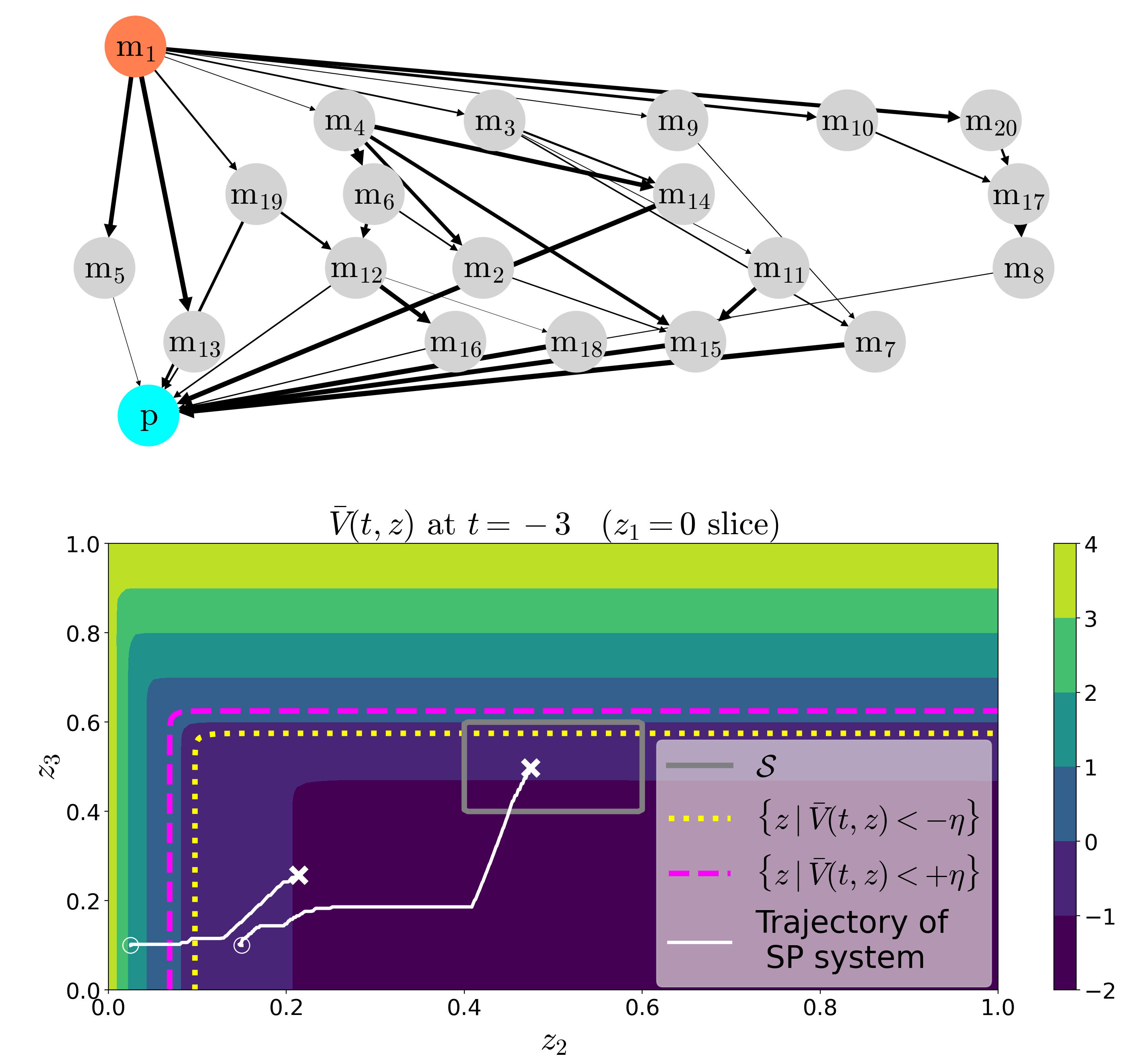}
    \caption{(Top) Example MRN, here with $N = 20$.
    Nodes in the network represent metabolites, and edges represent reactions.
    The upstream molecule is $\text{m}_1$ (coral), which is rapidly converted in metabolites that are themselves ultimately metabolized into the downstream molecule p (cyan).
    Edge thickness is proportional to the reaction's rate coefficient.
    If $T$ is the adjacency matrix of the graph, with $T_{ij}$ being the weight of the edge from node $j$ to $i$, one obtains the corresponding matrix $A_{\text{MRN}}$ in \eqref{eqn:big-system-fast} by taking the submatrix of $T$ corresponding to all nodes other than p and subtracting from each diagonal element the corresponding column sum.
    One obtains the corresponding matrix $C_{\text{MRN}}$ in \eqref{eqn:big-system-slow-1} by taking the row of $T$ corresponding to node p and subsequently eliminating the element of this row corresponding to node p.
    (Bottom) Contour plot of the value function $\valRM$ for the reduced model \eqref{eqn:big-system-rm-1}-\eqref{eqn:big-system-rm-3}, in which the metabolic network is as shown above.
    Here $t = -3$, $\uVals = [0,1]^2$, $\dVals = [0.9,1.1]$, and the weights of the edges in the metabolic network were sampled from a unit uniform distribution.
    Only the $z_1 = 0$ slice of the value function is shown.
    The target set $\mathcal{S}$ is the interior of the solid grey line, while the sub-level sets of the reduced value function for levels $\pm\eta = \pm0.5$ lie below the dashed magenta and dotted yellow lines, respectively.
    Trajectories of the SP dynamics \eqref{eqn:big-system-slow-1}-\eqref{eqn:big-system-slow-3} with $\ep = 0.01$ are shown as a white lines, with the initial state marked with an $\bigcirc$, and the final with an $\times$.
    The initial states are $\zSig_1(t) = 0$, $\zSig_2(t) = 0.025$, $\zSig_3(t) = 0.1$, $\ySig(t) = (0,\dots,0)$ and 
    $\zSig_1(t) = 0$, $\zSig_2(t) = 0.15$, $\zSig_3(t) = 0.1$, $\ySig(t) = (0,\dots,0)$.
    The optimal control law was chosen based on the value function computed for the reduced model, and the disturbance was chosen as a random signal of points uniformly sampled from $\dVals$.
    The target is $\targetSet = \R \times (.4,.6) \times (.4,.6)$ and the payoff function is $\ell(z) = \min\{\max\{10(z_2 - 0.5)-1, 10(z_3 - 0.5) - 1\}, 4 \}$.
    \label{fig:big-example-fig}}
\end{figure}

We next consider a model of a metabolic reaction network (MRN) in a growing population of cells.
In particular, a user supplies the population with a molecule $\text{m}_1$, which is rapidly metabolized by the cells into other intermediate metabolites $\text{m}_i$ ($i = 2\dots N$).
These intermediates are processed into a molecule p, which is required for cell growth.
Such an MRN is shown in Figure \ref{fig:big-example-fig} (Top) with $N = 20$.

Each cell additionally produces a separate molecule q, whose production rate can be also controlled (e.g. by optogenetics).
Ultimately, the user is interested in ensuring both the number of cells in the population and the amount of q produced are between some desired bounds at the final time.

Under the assumption that the rate of the metabolic reactions are much faster than the maximal rate of cell growth, we can consider the following model for this system:
\begin{align}
    \dot{\zSig}_1 &= C_{\text{MRN}} \dSig_1 \ySig - \frac{\zSig_1}{\zSig_1 + 1} \zSig_1 \label{eqn:big-system-slow-1}\\
    \dot{\zSig}_2 &= \lf(\frac{\zSig_1}{\zSig_1 + 1} - \zSig_2 \rg) \zSig_2 \label{eqn:big-system-slow-2}\\
    \dot{\zSig}_3 &= \zSig_2 \uSig_2 \label{eqn:big-system-slow-3}\\
    \ep \dot{\ySig} &= A_{\text{MRN}} \dSig_1 \ySig + \ha{e}_1 \uSig_1 \label{eqn:big-system-fast},
\end{align}
where $A_{\text{MRN}} \in \Ryy$, $C_{\text{MRN}} \in \R^{1 \tms \ny}$, and $\ha{e}_1 = \begin{pmatrix}
    1 & 0 & \dots & 0
\end{pmatrix}^\top \in \R^{\ny \tms 1}$.

In this model, the $i$-th entry of $\ySig$ is the concentration of molecule $\text{m}_i$ ($i = 1,\dots,N$), $\zSig_1$ is the concentration of molecule p, $\zSig_2$ is the population size, $\zSig_3$ is the total number of copies of molecule q in the population, $\uSig_1$ is the rate at which the user supplies $\text{m}_1$, $\uSig_2$ is the signal used to control production of $\text{q}$, and $\dSig_1$ is a possible disturbance to the population's global metabolic efficiency (with all quantities normalized).
Here, $A_{\text{MRN}}$ and $C_{\text{MRN}}$ are derived from the (weighted) adjacency matrix of the MRN (see caption of Figure \ref{fig:big-example-fig} for details), and $\epsilon$ is a parameter which is small when the metabolic reactions rates are much faster than the maximal cell growth rate.
Importantly, the matrix $A_{\text{MRN}}$ is invertible because all metabolites $\text{m}_1$, ..., $\text{m}_N$ are eventually converted into molecule p.

The reduced model for the above SP system is 
\begin{align}
    \dot{\zSig}_1 &= C_{\text{MRN}} A_{\text{MRN}}^{-1} \ha{e}_1 \uSig_1  - \frac{\zSig_1}{\zSig_1 + 1} \zSig_1 \label{eqn:big-system-rm-1}\\
    \dot{\zSig}_2 &= \lf(\frac{\zSig_1}{\zSig_1 + 1} - \zSig_2 \rg) \zSig_2 \label{eqn:big-system-rm-2}\\
    \dot{\zSig}_3 &= \zSig_2 \uSig_2 \label{eqn:big-system-rm-3}
\end{align}
\begin{remark}
    The natural domain for \eqref{eqn:big-system-slow-1}-\eqref{eqn:big-system-fast} is $(\zSig_1(\ta),\zSig_2(\ta)\zSig_3(\ta),\ySig(\ta)) \in [0,\ii) \tms [0,1] \tms [0,\ii) \tms \R^N$
    However, the results in this paper are for systems defined on all of $\R^{\nz + \ny}$.
    Such a discrepancy is of no concern because these dynamics can be extended to all of $\R^{\nz + \ny}$ in such a way that Assumption \ref{assumption:regularity} is satisfied.
\end{remark}

Level sets of the reduced value function are shown in Figure \ref{fig:big-example-fig} (Bottom).
While it is generally intractable to compute the value function for a model with 23 states (as in the SP model) for ground-truth comparison, we can demonstrate the utility of our method by using the reduced value function to provide a ``best guess" for the optimal control:
    $u^*(z,y,t) = \argmin_{\uVal \in \uVals} \max_{\dVal \in \dVals}
    \nabla_z \valRM(z,t)^\top F(z,u,d),$
where $F$ is the dynamics function of \eqref{eqn:big-system-rm-1}-\eqref{eqn:big-system-rm-3} [note that the above control is $y$-independent].
Indeed, by using this optimal control law estimate, we observe that two simulated trajectories of the SP model \eqref{eqn:big-system-slow-1}-\eqref{eqn:big-system-fast}, do and do not reach the target in accordance with the bounds in Theorem \ref{theorem:main-theorem}.

\section{CONCLUSION}
%auto-ignore
In this work, we identified a class of SP differential games that can be readily reduced by leveraging the results in \cite{gaitsgory}, and we related the BRS, BRT, and BST of the SP differential game to the reduced value function.
We then demonstrated the particular applicability of these results in analyzing biological systems, which often have dynamics that fit the identified class of systems.
In the second example, we empirically found that using the reduced value function to obtain an optimal control law allowed one to properly control the state of the SP model.
Future work may include studying when we can guarantee this procedure will well approximate the optimal control law of the SP differential game when the parameter $\epsilon$ is sufficiently small.
\appendix

For convenience throughout this appendix,
we will let $F: \Rz \tms \uVals \tms \dVals \to \Rz$ and $H: \Rz \tms \Rz \to \R$ be given by
\begin{align*}
    &F(\zVal,\uVal,\dVal) = f(\zVal,\uVal,\dVal) - M(\zVal) g(\zVal,\uVal,\dVal),\\
    &H(\zVal, \lm) = \min_{\uVal \in \uVals} \max_{\dVal \in \dVals} \lm^\top F(\zVal, \uVal, \dVal) = \max_{\dVal \in \dVals} \min_{\uVal \in \uVals}  \lm^\top F(\zVal, \uVal, \dVal).
\end{align*}

Given $t_0,\dots,t_n \in \R$ such that $t_0 < \dots < t_n$, and given functions $\aSig_1:[t_0,t_1) \to \R^m, \dots, \aSig_n: [t_{n-1},t_n) \to \R^m$, we define the function $[\aSig_1,\dots,\aSig_n):[t_0,t_n) \to \R^m$ such that
$[\aSig_1,\dots,\aSig_n)(s) =  \aSig_j(s)$ whenever $s \in [t_{j-1},t_j)$.
For each $[t,s) \ssbs \R$, we also let $\uSigsts$ $(\dSigsts)$ be the set of measurable functions from $[\tVal,s)$ to $\uVals$ $([\tVal,s)$ to $\dVals)$.

\subsection{Boundary Layer Model}
For each $\zVal \in \Rz$, we consider the following ``boundary-layer'' model:
\begin{equation}\label{eqn:blm}
    \dot{\ySig} = f(\zVal, \uSig, \dSig) + A(\zVal, \uSig, \dSig) \ySig.
\end{equation}

For each $\tVal < 0$, $\yVal \in \Ry$, $\zVal \in \Rz$, $\uSig \in \uSigs$, and $\dSig \in \dSigs$, we let $\trajBLM:[\tVal,0] \to \Ry$ be the unique Carath\'{e}odory solution of \eqref{eqn:blm} satisfying $\trajBLM(\tVal) = \yVal$.

For each $\tVal < 0$, $\yVal \in \Ry$, $\zVal \in \Rz$, and $\lm \in \Rz$, we define the payoff functional $\payoffBLM: \uSigs \tms \dSigs \to \R$ by
\begin{align*}
    \payoffBLM(\uSig,\dSig) = \frac{1}{|t|} \int_t^0 \lm^\top [ &f(\zVal, \uSig(\ta), \dSig(\ta)) +\\
    &M(\zVal)A(\zVal, \uSig(\ta), \dSig(\ta)) \trajBLM(\ta)] \dee \ta.
\end{align*}

Moreover, for each $\tVal < 0$, $\yVal \in \Ry$, $\zVal \in \Rz$, and $\lm \in \Rz$, and for each partition $P$ of the interval $[t,0)$ into the subintervals $[t_0,t_1),\dots,[t_{r-1},t_r)$ (where $t_0 = \tVal$ and $t_{r} = 0$), we define the following upper and lower value functions for the boundary-layer model:
\begin{align}
    \valBLMdiscretePlus(\tVal, \yVal) = \inf_{\uSig_1 \in \uSigsi{1}}  &\sup_{\dSig_1 \in \dSigsi{1}} \dots \inf_{\uSig_r \in \uSigsi{r}} \sup_{\dSig_r \in \dSigsi{r}} \nonumber\\
    &\payoffBLM([\uSig_1,\dots,\uSig_r), [\dSig_1,\dots,\dSig_r)), \label{eqn:af-game-value-plus}\\
    \valBLMdiscreteMinus(\tVal, \yVal) =   \sup_{\dSig_1 \in \dSigsi{1}} &\inf_{\uSig_1 \in \uSigsi{1}} \dots \sup_{\dSig_r \in \dSigsi{r}} \inf_{\uSig_r \in \uSigsi{r}} \nonumber\\
    &\payoffBLM([\uSig_1,\dots,\uSig_r), [\dSig_1,\dots,\dSig_r)), \label{eqn:af-game-value-minus}
    \end{align}
    where $\uSigsi{j} = \uSigsii{j-1}{j}$ and $\dSigsi{j} = \dSigsii{j-1}{j}$ for each $j$.

\subsection{Intermediate Lemmas}
\begin{lemma} \label{lemma:exponential-bounding}
    Let $\zVal \in \Rz$.
    There are constants $\al,\kp > 0$ such that for each $\tVal < 0$, $\uSig \in \uSigs$, and $\dSig \in \dSigs$,
    \begin{equation}
        \lf\| e^{\int_\tVal^0 A(z,\uSig(r),\dSig(r)) \dee r} \rg\| \le \al e^{\kp \tVal}.
    \end{equation}
\end{lemma}
\begin{proof}
    Let $\Lm: \uVals \tms \dVals \to \R$ be such that $\Lm(\uVal,\dVal)$ is the largest eigenvalue of $A(\zVal,\uVal,\dVal)^\top P + P A(\zVal,\uVal,\dVal)$, where $P$ is as in Assumption \ref{assumption:stability}.
    Then $\Lm$ is continuous, so by Assumption \ref{assumption:stability}, there is some $\nu > 0$ such that $\Lm(\uVal,\dVal) \le -\nu$ for each $\uVal \in \uVals$ and $\dVal \in \dVals$.

    Fix some $\tVal < 0$, $\uSig \in \uSigs$, and $\dSig \in \dSigs$.
    Let $\wSig:[\tVal,0] \to \Ry$ be the Carath\'{e}odory solution to 
    $$\dot{\wSig} = A(\zVal,\uSig, \dSig) \wSig, \quad \wSig(t) = \wVal.$$
    A standard argument (see e.g. the proof of Theorem 4.10 in \cite{Khalil}) using the Lyapunov function $V(\wVal) = \wVal^\top P \wVal$ shows
    \begin{equation*}
        \frac{\dee}{\dee \ta} [V(\wSig(\ta))] = - 2 \kp V(\wSig(\ta)) - \mu(\ta)
    \end{equation*}
    for a.e. $\ta \in (t,0)$, where $\kp := \frac{\nu}{2\lm_{\max}(P)}$ and $\mu(\cdot) \ge 0$.

    Since $\wSig$ is absolutely continuous with a compact domain, and since $V$ is locally Lipschitz, then $V \circ \wSig$ is absolutely continuous, so that
    $$V(\wSig(0)) = V(\wVal) e^{2 \kp \tVal} - \int_\tVal^0 \mu(s) e^{2 \kp s} \dee s \le V(\wVal) e^{2 \kp \tVal},$$
    and thus $\|\wSig(\tVal)\| \le \|\wVal\| \al e^{\kp \tVal}$,
    where $\al = \sqrt{\frac{\lm_{\max}(P)}{\lm_{\min}(P)}}$ (again see the proof of Theorem 4.10 in \cite{Khalil} for details).
    But since $\wVal \in \Ry$ was arbitrary and
    $\wSig(\tVal) = e^{\int_\tVal^0 A(\zVal,\uSig(r),\dSig(r))} \wVal$,
    the result follows.
\end{proof}

\begin{lemma} \label{lemma:bounding-payoff-error}
    Let $\yVal \in \Ry$, $\zVal \in \Rz$, and $\lm \in \Rz$.
    There is a $\bt > 0$ such that for all $\tVal < 0$, $\uSig \in \uSigs$, and $\dSig \in \dSigs$,
    \begin{align*}
        \lf|\payoffBLM(\uSig,\dSig) - \frac{1}{|\tVal|} \int_\tVal^0 \lm^\top F(\zVal,\uSig(\ta),\dSig(\ta)) \dee \ta\rg| \le \frac{\bt}{|\tVal|}.
    \end{align*}
\end{lemma}
\begin{proof}
    Fix some $\tVal < 0$ and some $\uSig \in \uSigs$ and $\dSig \in \dSigs$.
    Then $\payoffBLM(\uSig,\dSig) = I_1 + I_2 + I_3$, where
    \begin{alignat}{2}
        I_1 &= \frac{1}{|\tVal|} \int_\tVal^0 \lm^\top &&f(\zVal,\uSig(\ta),\dSig(\ta)) \dee \ta,\\
        I_2 &= \frac{1}{|\tVal|} \int_\tVal^0 \int_\tVal^\ta &&\lm^\top M(\zVal)A(\zVal,\uSig(\ta), \dSig(\ta)) e^{\int_\ta^0 A(\zVal,\uSig(r),\dSig(r)) \dee r}\nonumber \label{proof:lemma-bounding-payoff-I2}\\
        & &&g(\zVal,\uSig(s),\dSig(s)) \dee s \dee \ta,\\
        I_3 &= \frac{1}{|\tVal|} \int_\tVal^0 \lm^\top  && e^{\int_\ta^0 A(\zVal,\uSig(r),\dSig(r)) \dee r} \yVal \dee \ta.
    \end{alignat}
    Choose $\al > 0$ and $\kp > 0$ as in Lemma \ref{lemma:exponential-bounding}.
    Then
    \begin{equation*}
        |I_3| \le \frac{\al \|\lm\|\|\yVal\|}{|\tVal|} \int_\tVal^0 e^{\kp \ta} \dee \ta 
        = \frac{\al \|\lm\|\|\yVal\|}{\kp |\tVal|} \lf(1- e^{\kp \tVal} \rg).
    \end{equation*}
    By The Fundamental Theorem of Calculus for Lebesgue Integrals (see Theorem 3.35 in \cite{Folland}), for a.e. $\ta \in (\tVal,0)$,
    $$\frac{\dee}{\dee \tau} \int_\ta^0 A(\zVal,\uSig(r),\dSig(r)) \dee r = -A(\zVal,\uSig(\ta),\dSig(\ta)),$$
    so that
    $$\frac{\dee}{\dee \ta} e^{\int_\ta^0 A(\zVal,\uSig(r),\dSig(r)) \dee r} = -A(\zVal,\uSig(\ta),\dSig(\ta)) e^{\int_\ta^0 A(\zVal,\uSig(r),\dSig(r))}.$$
    Since $\int_\ta^0 A(\zVal,\uSig(r),\dSig(r)) \dee r$ is absolutely continuous as a function of $\ta$ on $[\tVal,0]$, and since the exponential of an absolutely continuous function on a compact interval is absolutely continuous, then $e^{\int_\ta^0 A(\zVal,\uSig(r),\dSig(r))}$ is also absolutely continuous as a function of $\ta$ on the interval $[\tVal,0]$. 
    Again by the The Fundamental Theorem of Calculus for Lebesgue Integrals, for each $s \in [\tVal,0]$
    \begin{align*}
        &e^{\int_s^0 A(\zVal,\uSig(r),\dSig(r)) \dee r} - I =\\
        &\int_s^0 A(\zVal,\uSig(\ta), \dSig(\ta)) e^{\int_\ta^0 A(\zVal,\uSig(r),\dSig(r)) \dee r} \dee \ta,
    \end{align*}
    where $I$ is the identity matrix in $\Ryy$.
    By changing the order of integration in \eqref{proof:lemma-bounding-payoff-I2},
    \begin{alignat*}{2}
        I_2 &= \frac{1}{|\tVal|} \int_\tVal^0 \int_s^0 &&\lm^\top M(\zVal)A(\zVal,\uSig(\ta), \dSig(\ta)) e^{\int_\ta^0 A(\zVal,\uSig(r),\dSig(r)) \dee r}\\
        & && g(\zVal,\uSig(\ta),\dSig(\ta)) \dee \ta \dee s\\
        &= \frac{1}{|\tVal|} \int_t^0 \lm^\top && M(\zVal) \lf[e^{\int_s^0 A(\zVal,\uSig(r),\dSig(r)) \dee r} - I\rg]\\
        & && g(\zVal,\uSig(s),\dSig(s)) \dee s.
    \end{alignat*}
    Note that
    \begin{align*}
        &\lf|\frac{1}{|\tVal|} \int_\tVal^0 \lm^\top M(\zVal) e^{\int_s^0 A(\zVal,\uSig(r),\dSig(r)) \dee r} g(\zVal,\uSig(s),\dSig(s)) \dee s \rg| \le \\
        &\quad\frac{\al \|\lm\| \|M(z)\| B}{|\tVal|} \int_\tVal^0 e^{\kp s} \dee s = \frac{\al \|\lm\| \|M(z)\| B}{\kp |\tVal|} \lf(1 - e^{\kp \tVal} \rg),
    \end{align*}
    where $B := \max_{\uVal \in \uVals}  \max_{\dVal \in \dVals} \|g(\zVal, \uVal, \dVal)\|.$
    The lemma then follows with
    $\bt := \frac{\al \|\lm\|}{\kp} \lf(\|\yVal\| + B\|M(z)\| \rg).$
\end{proof}

\begin{lemma} \label{lemma:payoff-limit}
Given $\lm,z \in \Rz$ and $\tVal < 0$,
\begin{align*}
    &\inf_{\uSig \in \uSigs} \sup_{\dSig \in \dSigs} \frac{1}{|\tVal|} \int_\tVal^0 \lm^\top F(z,\uSig(\ta),\dSig(\ta)) \dee \ta\\
    &= \sup_{\dSig \in \dSigs} \inf_{\uSig \in \uSigs} \frac{1}{|\tVal|}\int_\tVal^0 \lm^\top F(z,\uSig(\ta),\dSig(\ta)) \dee \ta = H(\zVal,\lm).
\end{align*}
\end{lemma}
\begin{proof}
    Let $\uVal^* \in \argmin_{\uVal \in \uVals}\max_{\dVal \in \dVals} \lm^\top F(\zVal,\uVal,\dVal)$, and let $\dVal^* \in \argmax_{\dVal \in \dVals}\min_{\uVal \in \uVals} \lm^\top F(\zVal,\uVal,\dVal)$.
    Then
    \begin{equation}\label{proof:lemma-payoff-limit-1}
      H(\zVal, \lm) = \max_{\dVal \in \dVals} \lm^\top F(\zVal, \uVal^*, \dVal) = \min_{\uVal \in \uVals}  \lm^\top F(\zVal, \uVal, \dVal^*)  
    \end{equation}
     by Assumption \ref{assumption:saddle-point}. Observe that
    \begin{align*}
        \inf_{\uSig \in \uSigs} &\sup_{\dSig \in \dSigs} \frac{1}{|\tVal|} \int_\tVal^0 \lm^\top F(\zVal,\uSig(\ta),\dSig(\ta)) \dee \ta \\
        &\le \sup_{\dSig \in \dSigs} \frac{1}{|\tVal|} \int_\tVal^0 \lm^\top F(\zVal, \uVal^*,\dSig(\ta)) \dee \ta \\
        &= H(\zVal,\lm) = \inf_{\uSig \in \uSigs} \frac{1}{|\tVal|} \int_\tVal^0 \lm^\top F(\zVal,\uSig(\ta),\dVal^*) \dee \ta \\
        &\le \sup_{\dSig \in \dSigs} \inf_{\uSig \in \uSigs} \frac{1}{|\tVal|} \int_\tVal^0 \lm^\top F(z,\uSig(\ta),\dSig(\ta)) \dee \ta \\
        &\le \inf_{\uSig \in \uSigs} \sup_{\dSig \in \dSigs} \frac{1}{|\tVal|} \int_\tVal^0 \lm^\top F(z,\uSig(\ta),\dSig(\ta)) \dee \ta
    \end{align*}
    where both equalities are justified by \eqref{proof:lemma-payoff-limit-1}, and where the final inequality is justified by the standard minimax inequality.
\end{proof}

\begin{lemma} \label{lemma:bounding-value-error}
Let $\yVal \in \Ry$, $\zVal \in \Rz$, and $\lm \in \Rz$.
    There is some $\bt > 0$ such that for all $\tVal < 0$ and for each partition $P$ of the interval $[\tVal,0)$,
    \begin{equation*}
    \lf|\valBLMdiscretePlusMinus(\tVal,\yVal) - H(\zVal,\lm) \rg| \le \bt/|\tVal|.
    \end{equation*}
\end{lemma}
\begin{proof}
    Let $\uSigsi{1},\dots,\uSigsi{r}$ and $\dSigsi{1},\dots,\dSigsi{r}$ be as in the definitions of $\valBLMdiscretePlusMinus$.
    First note that it follows from repeated use of the standard minimax inequality that
    \begin{align*}
    &\sup_{\dSig \in \dSigs} \inf_{\uSig \in \uSigs} \frac{1}{|\tVal|} \int_\tVal^0 \lm^\top F(\zVal,\uSig(\ta),\dSig(\ta)) \dee \ta \\
    &= \sup_{\dSig_1 \in \dSigsi{1}} \dots \sup_{\dSig_r \in \dSigsi{r}}\inf_{\uSig_1 \in \uSigsi{1}} \dots \inf_{\uSig_r \in \uSigsi{r}}\\
    &\quad\quad \frac{1}{|\tVal|} \int_\tVal^0 \lm^\top F(\zVal,[\uSig_1,\dots,\uSig_r)(\ta),[\dSig_1,\dots,\dSig_r)(\ta)) \dee \ta\\
    &\le \inf_{\uSig_1 \in \uSigsi{1}} \sup_{\dSig_1 \in \dSigsi{1}} \dots \inf_{\uSig_r \in \uSigsi{r}} \sup_{\dSig_r \in \dSigsi{r}} \\
    &\quad\quad \frac{1}{|\tVal|} \int_\tVal^0 \lm^\top F(\zVal,[\uSig_1,\dots,\uSig_r)(\ta),[\dSig_1,\dots,\dSig_r)(\ta)) \dee \ta\\
    &\le \sup_{\dSig_1 \in \dSigsi{1}} \dots \sup_{\dSig_r \in \dSigsi{r}} \inf_{\uSig_1 \in \uSigsi{1}} \dots \inf_{\uSig_r \in \uSigsi{r}} \\
    &\quad\quad \frac{1}{|\tVal|} \int_\tVal^0 \lm^\top F(z,[\uSig_1,\dots,\uSig_r)(\ta),[\dSig_1,\dots,\dSig_r)(\ta)) \dee \ta\\
    &= \inf_{\uSig \in \uSigs} \sup_{\dSig \in \dSigs} \frac{1}{|\tVal|} \int_\tVal^0 \lm^\top F(\zVal,\uSig(\ta),\dSig(\ta)) \dee \ta.
    \end{align*}
    It then follows from Lemma \ref{lemma:payoff-limit} that
    \begin{align}\label{proof:value-error-1}
        &H(\zVal,\lm) = \\
        &\quad\inf_{\uSig_1 \in \uSigsi{1}} \sup_{\dSig_1 \in \dSigsi{1}} \dots \inf_{\uSig_r \in \uSigsi{r}} \sup_{\dSig_r \in \dSigsi{r}} \frac{1}{|\tVal|} \int_\tVal^0 \lm^\top F(\zVal,\uSig(\ta),\dSig(\ta)) \dee \ta. \nonumber
    \end{align}
    Then by \eqref{eqn:af-game-value-plus} and \eqref{proof:value-error-1} together with repeated use of the standard inequality for the absolute difference of suprema/infima
    \begin{align*}
        &|\valBLMdiscretePlusMinus(\tVal,\yVal) - H(\zVal,\lm)| \le \\
        &\sup_{\uSig_1 \in \uSigsi{1}} \sup_{\dSig_1 \in \dSigsi{1}} \dots \sup_{\uSig_r \in \uSigsi{r}} \sup_{\dSig_r \in \dSigsi{r}} \Big|\payoffBLM([\uSig_1, \dots \uSig_r),[\dSig_1,\dots,\dSig_r))\\ 
        &\quad- \frac{1}{|\tVal|} \int_\tVal^0 \lm^\top F(\zVal,[\uSig_1, \dots \uSig_r)(\ta),[\dSig_1,\dots,\dSig_r)(\ta)) \dee \ta\Big|
    \end{align*}
    The result follows from Lemma \ref{lemma:bounding-payoff-error} (the result for $\valBLMdiscreteMinus$ follows similarly).
\end{proof}

\subsection{Proof of Lemma \ref{lemma:main-lemma}}

Having shown Lemma \ref{lemma:bounding-value-error}, Lemma \ref{lemma:main-lemma} is now a mostly straightforward consequence of the main result in \cite{gaitsgory}:

\begin{proof}[Proof of Lemma \ref{lemma:main-lemma}]
    Applying the main result (Theorem 4.2) in \cite{gaitsgory} to our setting requires verifying a number of assumptions, which we do now (unfortunately the assumptions are too extensive to reproduce in this text).
    We note that Assumption 1 in \cite{gaitsgory} follows from Lemma \ref{lemma:bounding-value-error} in this text;
    Assumptions (A1)-(A3) in \cite{gaitsgory} follow from Assumption \ref{assumption:regularity} in this text; Assumptions (B1)-(B3) in \cite{gaitsgory} follow from Theorem 6.2, 7.1, and 7.2 in \cite{gaitsgory} together with Assumption \ref{assumption:stability} in this text (as this assumption justifies Assumptions (C1) and (C2) in \cite{gaitsgory}, which are needed in Theorems 6.1, 7.1, and 7.2).
    [Also note that by Lemma \ref{lemma:bounding-value-error}, the quantities $R_\dl^{up}$, $R_\dl^{lo}$, and $R$ in the hypotheses of Theorem 4.2 are all equal to the quantity $-H$ defined in this text.]
    
    By \cite{Evans-Diff-Games} and \cite{Ishii-uniqueness}, $\valRM$ is the unique continuous viscosity solution (in the sense described in Remark 4.1 of \cite{gaitsgory}) of the Hamilton-Jacobi PDE
    \begin{equation*}
        \begin{cases}
        -\partial_\tVal v(\tVal,\zVal) + H(\zVal, \nabla_\zVal v(\tVal, \zVal)) = 0, & (\tVal,\zVal) \in \Omega,\\
        v(\tVal,\zVal) = \ell(\zVal), & (\tVal,\zVal) \in \partial \Omega,
        \end{cases}
    \end{equation*}
    where $\Omega = (-\ii,0) \tms \Rz$ and $\partial \Omega = \{0\} \tms \Rz$.
    Having verified the above assumptions, Theorem 4.2 in \cite{gaitsgory} implies $\valSP(t,z,y) \to \valRM(t,z)$ as $\ep \to 0^+$ uniformly on compact sets in $(-\ii,0] \tms \Rz \tms \Ry$, which is our desired result.
\end{proof}

\begin{remark}
    We note two nuances in this proof which may catch the attention of the careful reader.
    First, the theory in \cite{gaitsgory} uses the Friedman definition of the value function (see \cite{Friedman}), whereas we use the Elliot-Kalton definitions for $\valSP$ and $\valRM$.
    These definitions are indeed equivalent in our setting, because they are both the unique viscosity solutions to a corresponding Hamilton-Jacobi-Isaacs PDE by \cite{Evans-Diff-Games}, \cite{Barron-Evans-Jensen}, \cite{Ishii-uniqueness}.
    Second, the assumptions made in \cite{Evans-Diff-Games} and \cite{Barron-Evans-Jensen} on boundedness of the system dynamics virtually never hold for systems \eqref{eqn:sp-system-slow}-\eqref{eqn:sp-system-fast} and \eqref{eqn:reduced-system}. Moreover, the definition of a viscosity solution used in these two early works postulates boundedness and uniform continuity of the solution, whereas the definition in \cite{gaitsgory} merely postulates continuity.
    It is known (and in fact even stated in \cite{Evans-Diff-Games}) that such strong assumptions are not required to characterize the value function as the unique viscosity solution of a Hamilton-Jacobi PDE.
    Indeed the ``existence'' proofs from \cite{Evans-Diff-Games} and \cite{Barron-Evans-Jensen} (i.e. that the value functions of a differential game is a [possibly unbounded and continuous but not uniformly so] viscosity solution of the corresponding Hamilton-Jacobi-Isaacs PDE) follow similarly under only Assumption \ref{assumption:regularity} and continuity of the terminal payoff function $\ell$.
    On the other hand, the uniqueness result follows under these weaker assumptions and less restrictive definition of a viscosity solution from Theorem 2.5 in \cite{Ishii-uniqueness}.
    Both these observations are not new (c.f. Sections III.3.2, VIII.1, and the end of VIII in the classic text \cite{Bardi}).
\end{remark}

\bibliographystyle{ieeetr}
\bibliography{references}

\end{document}